\documentclass[journal]{IEEEtran}

\usepackage{algorithm}
\usepackage{algorithmicx}
\usepackage{algpseudocode}
\usepackage{CJK}
\usepackage[top=2cm, bottom=2cm, left=2cm, right=2cm]{geometry}

\usepackage{amssymb}

\usepackage{subfigure} 
\usepackage{array}
\usepackage{amsmath}
\usepackage{amsmath,bm}
\usepackage{booktabs}
\usepackage{makecell}
\usepackage{multirow}
\usepackage{stfloats}
\usepackage{diagbox}
\usepackage{cite}
\usepackage[numbers,sort&compress]{natbib}
\usepackage{threeparttable}
\usepackage{varwidth}
\usepackage{stfloats}
\usepackage{graphicx}
\usepackage{soul}

\ifCLASSINFOpdf
\else
\fi
\begin{document}

%
\title{Continuous Prediction of Lower-Limb Kinematics From  Multi-Modal Biomedical Signals}
%
%
%
\author{Chunzhi~Yi,
	   Feng~Jiang$^*$,
             Shengping Zhang,
             Hao~Guo,
             Chifu~Yang,
        Zhen~Ding,
        Baichun~Wei,
        Xiangyuan~Lan$^*$,
        and~Huiyu~Zhou
\thanks{Chunzhi Yi, Hao Guo, Zhen Ding and Chifu Yang are with the School
of Mechatronics Engineering, Harbin Institute of Technology, Harbin,
Heilongjiang, 150001 China e-mail: chunzhiyi@hit.edu.cn(C.Yi).}
\thanks{Chunzhi Yi, Hao Guo, Baichun Wei and Feng Jiang are with the School of Computer, Harbin Institute of Technology, Harbin, Heilongjiang, 150001 China and Pengcheng Laboratory,  Shenzhen, Guangdong, China e-mail: fjiang@hit.edu.cn(F.Jiang).}
\thanks{Shengping Zhang is with the School of Computer Science and Technology,
Harbin Institute of Technology, Weihai 264209, China.}
\thanks{Xiangyuan Lan is with the Department of Computer Science, Hong Kong Baptist University, Hong Kong email: xiangyuanlan@life.hkbu.edu.hk (Corresponding author: Feng Jiang, Xiangyuan Lan). }
\thanks{Huiyu Zhou is with the School of Informatics, University of Leicester,
Leicester LE1 7RH, U.K.}}

%
%

\markboth{IEEE Transactions on Circuits and Systems for Video Technology,~Vol.~X, No.~X, Month~Year}%
{Shell \MakeLowercase{\textit{et al.}}: Bare Demo of IEEEtran.cls for IEEE Journals}
%



\maketitle

\begin{abstract}
The fast-growing techniques of measuring and fusing multi-modal biomedical signals enable advanced motor intent decoding schemes of lower-limb exoskeletons, meeting the increasing demand for rehabilitative or assistive applications of take-home healthcare. Challenges of exoskeletons’ motor intent decoding schemes remain in making a continuous prediction to compensate for the hysteretic response caused by mechanical transmission.  In this paper, we solve this problem by proposing an ahead-of-time continuous prediction of lower-limb kinematics, with the prediction of knee angles during level walking as a case study.
Firstly, an end-to-end kinematics prediction network(KinPreNet)\footnote{The demo is avalable on https://youtu.be/CMwtrVaeLZ4, and the source
code is attached to the IEEE Code Ocean.}, consisting of a feature extractor and an angle predictor, is proposed and experimentally compared with features and methods traditionally used in ahead-of-time prediction of gait phases. Secondly, inspired by the electromechanical delay(EMD), we further explore our algorithm's capability of compensating response delay of mechanical transmission by validating the performance of the different sections of prediction time. And we experimentally reveal the time boundary of compensating the hysteretic response.  Thirdly, a comparison of employing EMG signals or not is performed to reveal the EMG and kinematic signals' collaborated contributions to the continuous prediction. During the experiments, EMG signals of nine muscles and knee angles calculated from inertial measurement unit (IMU) signals are recorded from ten healthy subjects. Our algorithm can predict knee angles with the averaged RMSE of 3.98 deg which is better than the 15.95-deg averaged RMSE of utilizing the traditional methods of ahead-of-time prediction. The best prediction time is in the  interval of 27ms and 108ms. To the best of our knowledge, this is the first study of continuously predicting lower-limb kinematics in an ahead-of-time manner based on the electromechanical delay (EMD).
\end{abstract}


%
\IEEEpeerreviewmaketitle

\section{Introduction}
%
%
%
%
\IEEEPARstart{T}HE increasing number of gait-impaired patients, led by recent advancement in medical treatment and expanded life expectancy \cite{EMERY199119,doi:10.1111/dmcn.12826}, casts a rising demand on rehabilitative and assistive devices for take-home healthcare \cite{martini2019gait,voluntary,softrob}. Lower-limb exoskeletons, functioned as a home-healthcare device for both rehabilitation and assistive applications, have been enabled by the fast-growing techniques of measuring and fusing multi-modal biomedical signals  \cite{voluntary, garate2016walking}.  Fusing multi-modal biomedical signals sought to decode human motor intent, i.e. to perceive subject-specific gait characteristics, thus can result in improved assistive and/or rehabilitative performance of assistance.  Decoding human motor intent, which highly relies on kinematics-related information, is still a challenging topic of exoskeletons especially for assisting subjects still with mobility.  Traditionally, pressure
insoles allow the robot to automatically detect users'
gait phases according to the different pressure distribution
under feet during one gait cycle. However, such movement
intent decoding might result in a severe intent misjudgment
and thus a potential injury if stepping on a protuberance,
while cumbersome customized shoes have to be embedded
into the robot \cite{razak2012foot}. Alternatively,  some movement intent decoding schemes utilized kinematics-related characteristics to detect key timings of gait \cite{ mooney2014autonomous, 7394183, ding2016effect}, demonstrating the superiority of employing kinematics information. However, on one hand, the key timings provide discrete information of gait and movement intent, thus  might limit the further improvement of assistive performances. On the other hand, as revealed by the simulation study \cite{fang2019simulation}, the response delay caused by the transmission of the mechanical structure of exoskeletons will impede the close coordination between human and exoskeleton and thus greatly degrade the assistive performance. How to compensate the transmission delay is ignored by currently existing control and movement intent decoding schemes of lower-limb exoskeletons.

The acquisition of continuous kinematics could be enabled by techniques like adaptive oscillator (AO) or Bayesian filter. AO  \cite{chen2016gait, garate2016walking} learns periodic characteristics of locomotion by extracting frequency components from kinematics and then using the components to reconstruct the reference curves. According to such reference curves, AO estimates the gait percentage of stride, the information of which is limited to reflect kinematics in detail.  In \cite{8886526}, iterative extended Kalman filter was employed to make a one-step-ahead prediction of richer gait-related information, including gait events and trajectories of lower-limb joints.  However, the methods in \cite{chen2016gait, garate2016walking,8886526}, based on learning from the previous reference curves, rendered strong dependence on the periodic characteristics of gait. Additionally, the one-step-ahead prediction might be too short to fully compensate for the response delay of the mechanical transmission. 

Neural signals, which are generated prior to the corresponding movements, provide a promising solution to the compensation of response delay. Myoelectric signals, one of the main sources for obtaining
neural activation have been widely used in the control of
exoskeletons \cite{ sankai2010hal}, powered prostheses \cite{huang2005gaussian,  amsuss2013self, chan2004continuous} and rehabilitation
robots \cite{van2010limb, grabowski2013effects, perry2006design, connolly2008prosthetic,liu2012novel, matsubara2013bilinear, li2013boosting}. In decoding EMG signals, a pioneer work of H. Huang \cite{huang2008strategy}, which investigated the EMG-based pattern recognition (PR) methods to identify subjects’ locomotion modes, demonstrated the feasibility of using EMG signals to make an ahead-of-time prediction. Following works like \cite{huang2011continuous} and \cite{naik2018ica} applied such EMG-based pattern recognition methods on different locomotion modes, which provided additional demonstrations. However, given that recent assistive strategies of lower-limb exoskeletons are developed to require continuous information of lower-limb kinematics, such PR-based methods just focus on the qualitative and discrete presentation of lower limb motions, making it difficult to provide further information. 

One solution to continuous EMG-to-kinematics mapping is the musculoskeletal model combined with dynamic models. A study on upper limbs \cite{han2015state}, which included musculoskeletal and dynamic models into Kalman filter, demonstrated the feasibility of calculating kinematics based on EMG signals. However, if we apply this method on lower limbs, the ground-feet wrench, which is intractable to estimate using wearable sensors, is unavoidably employed in dynamic models. This issue impedes musculoskeletal and dynamic models' combined usage on lower limbs, thus results in either time-costing offline parameter identification of musculoskeletal models \cite{7497599} or a real-time framework enabled by force plates \cite{7927483}. Alternatively, Brantley et al. \cite{brantley2017prediction} directly mapped EMG into knee angles using an unscented Kalman filter, which was limited by the drawbacks of the Kalman filter itself. All the above-mentioned EMG-to-movement mapping methods, albeit capable of continuously acquiring kinematics, did not demonstrate the capability of making ahead-of-time predictions. 

The goal of this study is to make a continuous prediction of kinematics ahead of time. Particularly, a case study is performed on knee angle prediction. We propose to explore and exploit the EMG's characteristics of its onset before the onset of its corresponding movements (denoted as electromechanical delay (EMD)) by the means of an end-to-end LSTM network, in order to make the prediction. The architecture of the algorithm is made comparisons with traditionally used EMG features and prediction algorithms. Furthermore, based on experimental validations, we explore the time boundary of how much transmission delay can be compensated by the continuous kinematics prediction and reveal how EMG and kinematics signals can contribute to the final prediction performance. The main contributions of this paper are summarized as follow:


\begin{itemize}
\item To the best of our knowledge, this is the first study of  continuously predicting lower-limb kinematics in an ahead-of-time manner based on the electromechanical delay (EMD).

\item We experimentally revealed EMG and kinematics signals' collaborated contribution to the continuous kinematics prediction.

\item  We further explore the EMD-inspired prediction time through different trials of prediction time and experimentally revealed our algorithm's time boundary of compensating mechanical response delay.
 
\end{itemize}

This paper has been
organized as follows. Related works are presented in Section II.
Section III details the methodology and experiments. Experimental
results have been explained in Section IV. Discussion
and conclusion of the entire research work have been given in
Sections V and VI, respectively.

\section{Related Work}
Challenges of constructing the algorithm include solving the continuous prediction, determining prediction time and extracting effective features of EMG signals.

\subsection{Continuous and Ahead-Of-Time Prediction}

The continuous and ahead-of-time prediction can be inspired by previous works from two aspects. Firstly, the ahead-of-time prediction of discrete gait phases, was demonstrated with feasibility by the pioneer works of Huang et. al. \cite{huang2008strategy}. In Huang's works, support vector machine (SVM) classifier was leveraged to continuously classify  locomotion modes and predict the transmission of locomotion modes ahead of time, which presented better performances than linear discriminant analysis (LDA). Secondly, methods of upper-limb movement regression demonstrated the robustness and accuracy of Recurrent Neural Network (RNN) on such kinematics regression tasks \cite{xia2018emg, wang2002prediction}. Due to the long-term discrepancy problem of RNN, Long Short Term Memory model (LSTM), which was developed for processing sequence information with multi-layer neuron cells \cite{10.1093/bioinformatics/btz111}, could be proper to decode the continuity correlation between non-static EMG signals and kinematics under noisy measurements. Thus, in this study, a comparison was made between LSTM and SVM.

\begin{figure*}[h]
\centering 
\includegraphics[width = 18cm]{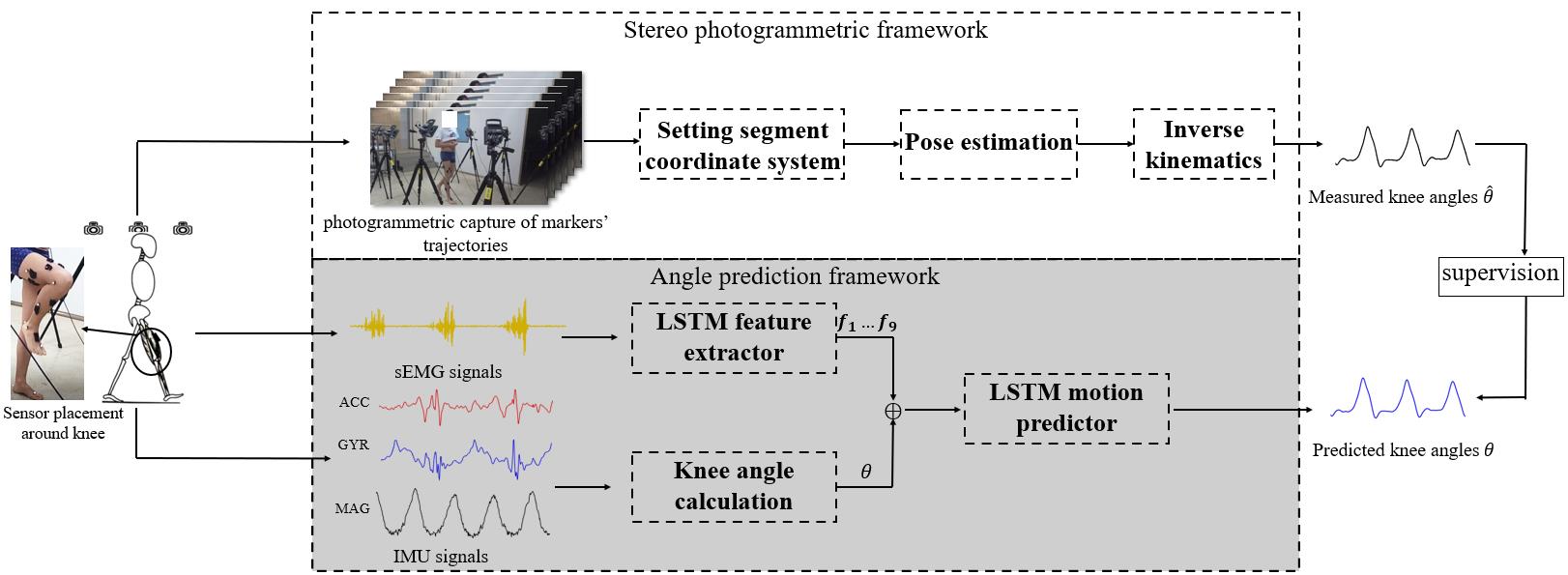}
\caption{Our teacher-student paradigm used in our method provided a cross-modal supervision for the . The upper pipeline provided a training supervision, whereas the bottom learned to predict joint angles using EMG and IMU signals.} 
\label{fig:workflow} 
\end{figure*}

\subsection{Prediction Time}
The prediction time, which is denoted by the interval between the timings of the prediction and the actual movement, is related to the data labelling of the algorithm, as will be shown in Section III-D. Determining the prediction time can give a precise reference on how much the mechanical delay can be compensated.  Huang et. al. \cite{huang2011continuous} proposed to make a prediction  based on  identifying the critical timing of two stable patterns' transition. That is, the proposed prediction time was related to the fixed consequence of gait phases. However, in this study, it might be difficult to leverage the discrete information of gait phases to perform the task of continuous kinematics prediction. Inspired by the fact that EMG signal is generated 25-125 ms before its corresponding actual motion  \cite{ blackburn2009comparison, LI2004647,emd2014}, denoted as the electromechanical delay (EMD), our algorithm utilizes this characteristic to perform continuous kinematics prediction and we further explore the maximum of transmission delay that can be compensated by the prediction time of the algorithm.

\subsection{Feature extraction of EMG signals}

Feature extraction of EMG signals is also a challenge for the continuous prediction due to the lack of features with demonstrated effectiveness and robustness on continuously predicting lower-limb kinematics. Huang et. al. made ahead-of-time prediction of discrete information using four time-domain features \cite{huang2011continuous}. Ngeo et. al.  \cite{ngeo2014continuous} investigated simultaneous and multiple finger kinematics estimation based on comparing four time-domain features and envelope of multi-channel surface EMG signals. The result showed that traditional EMG time-domain features outperformed filtered EMG envelop. Traditional time-domain features allow us to effectively recognize patterns and joint continuous parameters. However, they ignored the influence of EMG fluctuation and the antagonism of muscles, therefore cannot reflect the magnitude matchup between EMG signals and relative movements. Thus, in this paper, a feature extractor that could extract in-depth information of multi-channel EMG signals is enabled by the end-to-end training of the LSTM network and compared with traditionally used time-domain features.

\section{Methodology}

\subsection{Overview of the Proposed Model}

Our method, as  shown in Fig. \ref{fig:workflow}, follows a teacher-student paradigm. The top pipeline calculated joint angles using data from the stereo-photogrammetric framework, which provided a cross-modal supervision; the bottom pipeline predicted angles using IMU and EMG signals, which  performed an ahead-of-time prediction. To be specific, we developed an end-to-end LSTM network to predict kinematics with data labelled according to different prediction time. The end-to-end network was designed that it can be divided into feature extractor and motion predictor. Other than the LSTM feature extractor and LSTM motion predictor, we also employed traditionally-used time-domain features and SVM, in order to contribute to a comparison study.

\subsection{Cross-Modal Supervision}
The reason of employing the stereo-photogrammetric framework was to provide a "gold standard" of joint angles so as to have the angle prediction framework supervised with accurate labels. As shown in the top pipeline of Fig. \ref{fig:workflow}, the measured angles were obtained from videos. The measured 3-D locations of markers were firstly filtered by a 4th order Butterworth low-pass filter (cutoff frequency 6 Hz).  The coordinate system of each segment was set according to the definition of anatomical orientations and the placement of markers  \cite{DAVIS1991575, Jan2007ColorAO}. Particularly, the markers attached to analytical landmarks were recognized and tracked by the stereo-photogrammetric system. And the bone-embedded frames, i.e., the coordinate system of each segment, were determined according to its definition \cite{CAPPOZZO1995171} and at least three markers of each segment. In this way, a multi-link kinematics model can be built. Then, pose estimation was applied to extract positions and orientations of segments from markers  by the means of a global optimization process, which was reported with the advantage of minimizing the soft tissue and measurement error \cite{LU1999129,LEARDINI201777}. The pose estimation can be formulated as 
\begin{align}
\label{e2}
\min\limits_{q} (\sum\limits_{i}^{N} (\omega_i p_i^{measured} - T(q) p_i))
\end{align}
where $q$ denoted the generalized coordinates of the multi-link model, $\omega_i$ denoted the weight of the ith landmark, $T(q)$ denoted the transmission from the local segment-fixed coordinate frame to the laboratory frame, $p_i^{measured}$ denoted the measured position vector of the $i$th landmark with respect to the laboratory frame and  $p_i$ denoted the position vector of the $i$th landmark with respect to the segment-fixed coordinate frame. Finally, a 7-link inverse kinematics modeling of subjects (the conventional gait model) was performed after pose estimation, while anthropometric measurements of subjects were used to scale the model.  Joint angles were calculated using the joint axes of the inverse kinematics model and the generalized coordinates of the adjacent segments \cite{CAPPOZZO1995171,measurement}.

The stereo-photogrammetric framework took images as input and calculated joint angles as measured angles $\hat \theta$. The measured angles  $\hat \theta$ provided cross-modal supervision for the angle prediction framework, which learned to make an ahead-of-time prediction of angles as stated in the following sections. 

\subsection{Signal Preprocessing and Data Windowing}

\begin{figure}[t]
\centering 
\includegraphics[width = 6cm]{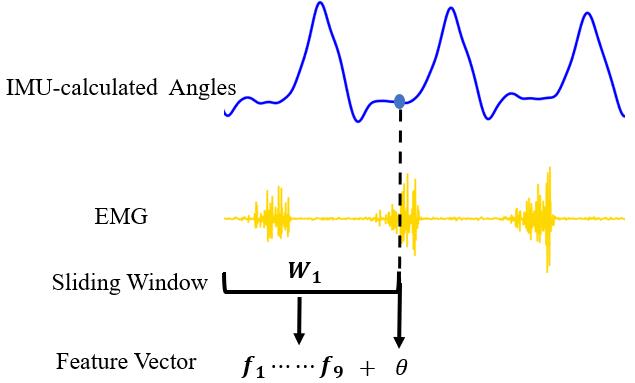}
\caption{Data windowing scheme. The feature vector was formed by combining EMG features of a sliding window and the current
IMU-calculated knee angle, where $\theta$ denoted the IMU-calculated knee angle.} 
\label{fig:03} 
\end{figure}

As presented in Fig.\ref{fig:workflow}, knee angles were firstly calculated by the method proposed in \cite{yi2018estimating} using IMU measurements. The calculation accuracy of the method was presented to be 1.67 deg during dynamic motions. Then, the calculated angles combined with EMG signals segmented by sliding windows were used to form feature vectors, as shown in Fig.\ref{fig:03}. To be specific, following the experience of literatures \cite{ngeo2014continuous, huang2011continuous}, the length of the sliding window was 148.5ms, and the increment was 13.5ms. The feature vector was constructed by combining EMG features extracted from nine channels with the IMU-calculated knee angle at the end
of the sliding window,
which is given by
\begin{align}
\label{e1}
\bm{x} = [\bm{f_1}, \bm{f_2}, \ldots, \bm{f_9}, \theta]
\end{align}
where $\bm{f_i}, i = 1,\cdots,9$ denoted the extracted features from each EMG channel, 
 $\theta$ denoted the calculated knee angles from IMU measurements.

\subsection{Feature Extraction}
Features in the time domain (FT):
Past studies have demonstrated the effectiveness of time-domain features traditionally used in PR-based prediction for
EMG-controlled lower-limb prostheses \cite{huang2011continuous, sutskever2014sequence}. In order to test
their performance in continuous motion prediction, we first
extracted such traditional features from EMG signals. Four traditionally-used
EMG time-domain features \cite{sutskever2014sequence} (mean absolute value, number
of slope sign changes, number of zero crossings and waveform
length) were simultaneously extracted in this experiment. As
presented in Eq. (\ref{e1}), EMG features, $\bm{f_n}$, each containing
four time-domain features were extracted from nine channels to construct a 37-element feature vector with a knee angle $\theta$.

Features from LSTM (FL):
Given that lower-limb movements are driven by the coordinative and antagonistic activations of muscles, FT, which was extracted from each single channel, cannot fully reflect the coordinative and antagonistic efforts of muscle activations, i.e. inhibition and excitation mechanisms \cite{reflex},  during level walking. Thus, information that represented such mechanism and/or muscle activation should be included in the extracted features in order to potentially improve the performance of the continuous prediction. However, currently used methods cannot perform this task well. The envelope
of EMG signals, although was mostly regarded as muscle activations,
cannot give a full insight of the coordinative and antagonistic efforts of muscles. To this end,  we employed an artificial neural network with the aim of  automatically extracting EMG features, expecting to break the potential bottleneck. Herein, LSTM, with the capability of forgetting useless information, jointly processed the EMG signals of all the channels, thus could effectively extract the inhibition and excitation mechanisms among EMG signals and highlight the correlation between joint efforts of muscle activations and movements.
Particularly, a four-layer LSTM was used to extract features from the nine-channel
EMG signals. The four-layer extractor consisted of three LSTM layers (40 as hidden size and 60 time steps) and a 40-by-9 fully connected layer. The topology was determined by multiple trials with the aim of realizing the simplest architecture while preserving enough accuracy.  In order to maximally remain the basic information into EMG signals, the input and output size
of the extractor was set to be equal. Then, the nine EMG features extracted by the feature extractor were combined with the calculated knee angles and concatenated into the 10-element feature vectors through Eq. (\ref{e1}), following the data windowing scheme presented in Fig. \ref{fig:03}. During the training session, both the LSTM predictor and the LSTM extractor were trained together as an end-to-end network. During the working session, we employed  the well-trained feature extractor to work with other predictors.

Features from LSTM and Time domain (FLT): When a
comparison was made between FT and FL, one question of
interest was whether they incorporate complementary
information to each other. If so, a combination of FT and FL
could provide more global insight of EMG signals.
To answer this question, FT and FL were combined together,
then used to construct a 46-element feature vector for testing
their performance on each predictor.

\subsection{Angle Predictor}
Due to its successful usage on processing  sequence signal
in translation \cite{sutskever2014sequence}, LSTM was employed to perform  the many-to-one mapping between EMG and its consequent movement.  
A 5-layer LSTM(40-40-40-80-1, the last two layers of which were fully connected layers), which
incorporated a forget gate and remember gate, was utilized as a predictor. Particularly, the 5-layer predictor was consisted of three LSTM layers (40 as hidden size and 60 as time step) and two fully connected layers with the size of 40*80 and 80 *1. 

Following SVM's demonstrated performance in predicting locomotion modes and real-time application of EMG-controlled hand prostheses, SVM was selected to contribute to a comparison study. The regression version of SVM, support vector regression (SVR) was employed as the other angle predictor. The applied kernel function was the  radial basis kernel function(RBF).

\subsection{Labeling and Prediction Time}
\begin{figure}[t]
\centering
\includegraphics[width = 6.5cm]{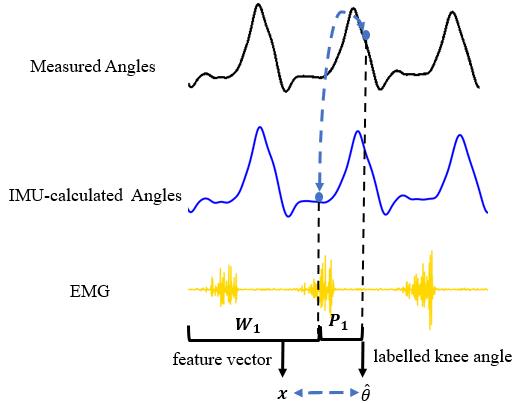}
\caption{Data labeling. Feature vectors were labelled by angles measured by the stereo-photogrammetric framework. The prediction time was set according to EMD. $\hat\theta$ denoted the angle measured by the framework} 
\label{fig:05} 
\end{figure}
As shown in Fig.\ref{fig:05}, each feature vector was labeled with the knee angle $\hat{\theta}$ measured by the stereo-photogrammetric framework after prediction time. The length of the prediction time $P_1$ was set inspired by the physiological time interval of EMD. 

To evaluate the influence of prediction time and to explore the time boundary,  six time sections  (27 ms, 54 ms, 81 ms,  108 ms, 135ms and 162ms) were selected considering the requirement of data synchronization and the inspiration of the normal
physiological time interval of EMD. 
All data were collected and labeled before the training session.
Cure parameters of each well-trained model, including
the LSTM extractor, the SVM predictor and the LSTM predictor, varied
with different sets of prediction time.

\section{Experiments}
\label{sec_result}
\subsection{Experimental Setup and Design}
\begin{figure}[h]
\centering 
\includegraphics[width = 6cm]{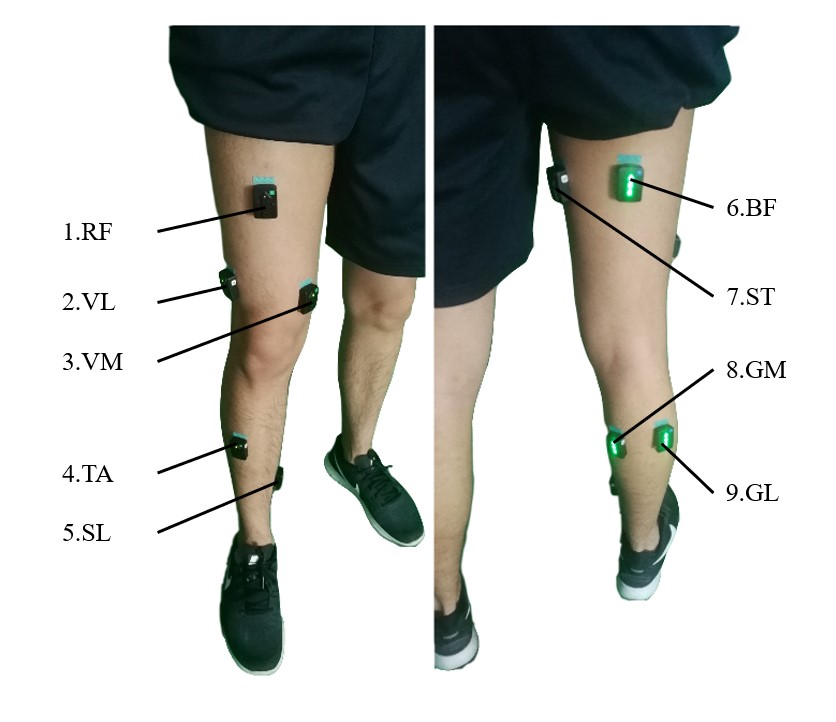}
\caption{Selected muscles and sensor attachment on subjects.}
\label{fig:01} 
\end{figure}

Ten healthy subjects (eight men and two women, age = 25$\pm$5 years, height = 1.75$\pm$0.05 m, weight = 67.6$\pm$12
kg) are  asked to walk with self-selected speeds. As shown in Fig. \ref{fig:01}, nine muscles from one leg were selected to incorporate most functional muscles relative to normal walking,
including: rectus femoris (RF), vastus lateralis muscle (VL),
vastus medialis muscle (VM), tibialis anterior muscle (TA),
soleus (SL), biceps femoris muscle (BF), semitendinosus (ST),
gastrocnemius muscle medial head (GM) and gastrocnemius
muscle lateral head (GL).

Surface electrodes (Delsys Trigno, IM type $\&$ Avanti type)
are placed on target muscles through palpation and skin preparation. In addition,  data from nine-axis inertial measurement units (IMUs)  is provided by the inertial sensors embedded in electrodes.  Sixteen retro-reflective markers are attached to subjects' pelvis and lower limbs. The markers are attached to the analytical landmarks of body \cite{LEARDINI2011562} according to the instructions and principles of \cite{LEARDINI2011562,Jan2007ColorAO}, and guaranteed by palpation \cite{huang2008strategy}.  The  3-D locations of the markers are recorded (100 Hz) using
a 8-camera vedio system (Vicon, Oxford, UK). The signals from EMG sensors (1111.11 Hz), IMUs (74Hz) and the vedio system  (100 Hz) are recorded and synchronized by time stamps.

In the experiment,  three minutes of standing still are provided for
initializing the joint angle calculation method. Five trials are performed on each subject, each trial lasts at least three minutes. Rest periods are allowed between trials to avoid fatigue. Before starting the experiment, anthropometric measurements like the height, weight and lengths of each lower-limb segments are measured in order to scale the model of the stereo-photogrammetric framework. 

The experiment protocol is approved by Chinese Ethics Committee of Registering Clinical Trials and all participants have been informed of the
content and their right to withdraw from the study at any
time, without giving any explanation.

\subsection{Implementation Details}

The training parameters, i.e. batch size, optimizer, Epoch, and the learning rate of extractor and predictor is set as batch size= 1, optimizer=Adam, Epoch=30, the initial learning rate of extractor $\&$ predcitor=0.001 and 0.0001, respectively. Every 20 times of training, the learning rates reduce 20$\%$.

Experiments are constructed to test the performance of different motion predictors (P),
extracted features sets (F) and prediction time (T), during which
each set of the three indicators are evaluated. During such
experiments, 10-fold cross-validation is applied to evaluate
our algorithms without the loss of generality. In the cross-validation procedure, data of a subject are used as the
testing database, while data of the remaining subjects are
used as the training database. This procedure is repeated for
each subject so that data from each subject could be used as
a testing database once. All the data collected during experiments form the dataset, which  consists of over 90,000,000 sample points of EMG signals, over 670,000 sample points of IMU-calculated knee angles and over 900,000 sample points of measured knee angles. A comparison is also made to evaluate the prediction’s reliance on periodic gait characteristics versus EMG, in order to reveal the collaborated contribution of EMG and kinematic signals.

With the well-trained LSTM extractor, SVR predictor and LSTM
predictor, the predicted knee angles are analyzed to evaluate the performance of different sets of (P, F, T). In order to distinguish the separate influence of the extracted features (F) and
prediction time (T), we present our results by averaging each of them, which is
\begin{align}
I_F = \frac{\sum_T I_{F,T}}{n_T}, I_T = \frac{\sum_F I_{F,T}}{n_F}
\end{align}
where $I$ denotes any evaluation index (e.g. RMSE, SNR, R-value or adjusted R), F denotes a feature extractor, T denotes a time section and $n_T$ ,$n_F$ denote the amount
of time sections and the number of feature extractors, which is equal
to 4 and 3 respectively.  The one-way ANOVA is performed on the
results to depict the repeated measures analysis of variance. A significant level is set to $\alpha$ = 0.05 for all testing.

\begin{figure*}[htbp] 
\centering 
\begin{minipage}[c]{0.15\textwidth} 
\centerline{\includegraphics[width = 5.3cm]{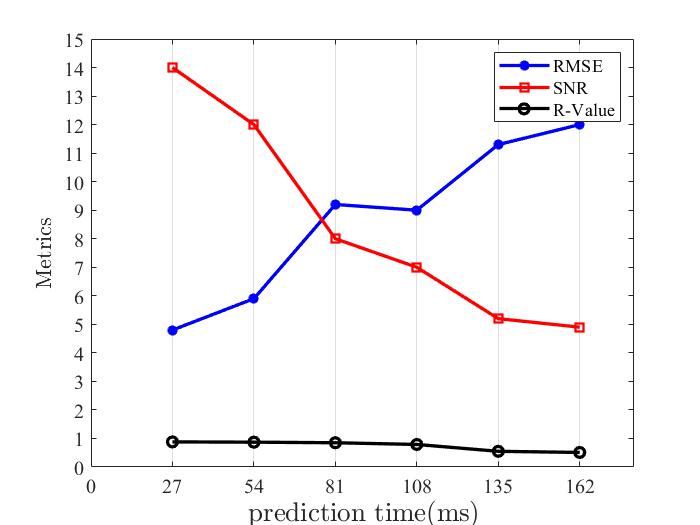}} 
\caption{The performance plots of different prediction time.} 
\label{fig:061} 
\end{minipage} 
\hfill 
\begin{minipage}[c]{0.82\textwidth} 
\centering 
\subfigure[RMSE]{
\includegraphics[width=4.5cm]{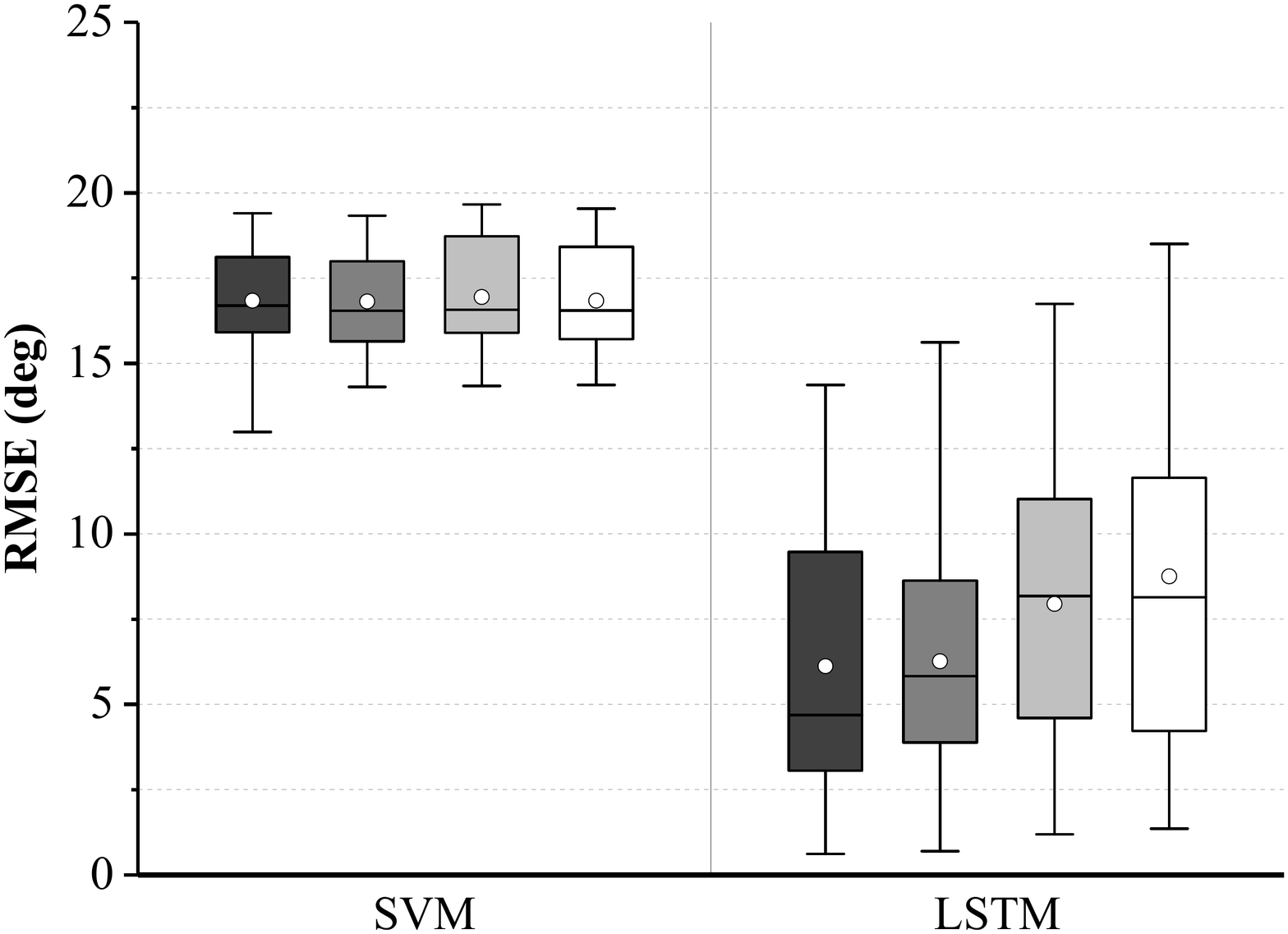}}
\subfigure[SNR]{
\includegraphics[width=4.5cm]{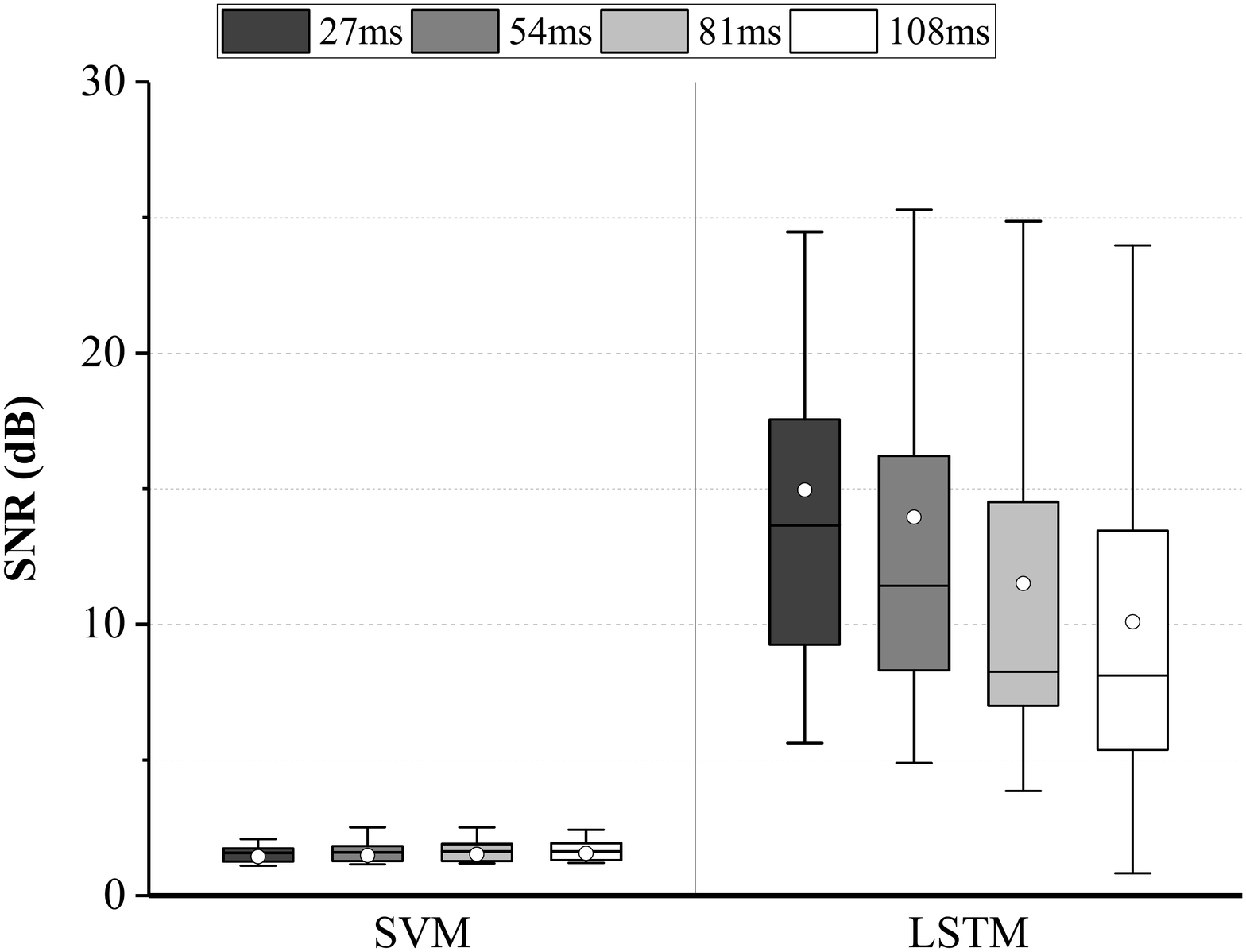}}
\subfigure[R-value]{
\includegraphics[width=4.5cm]{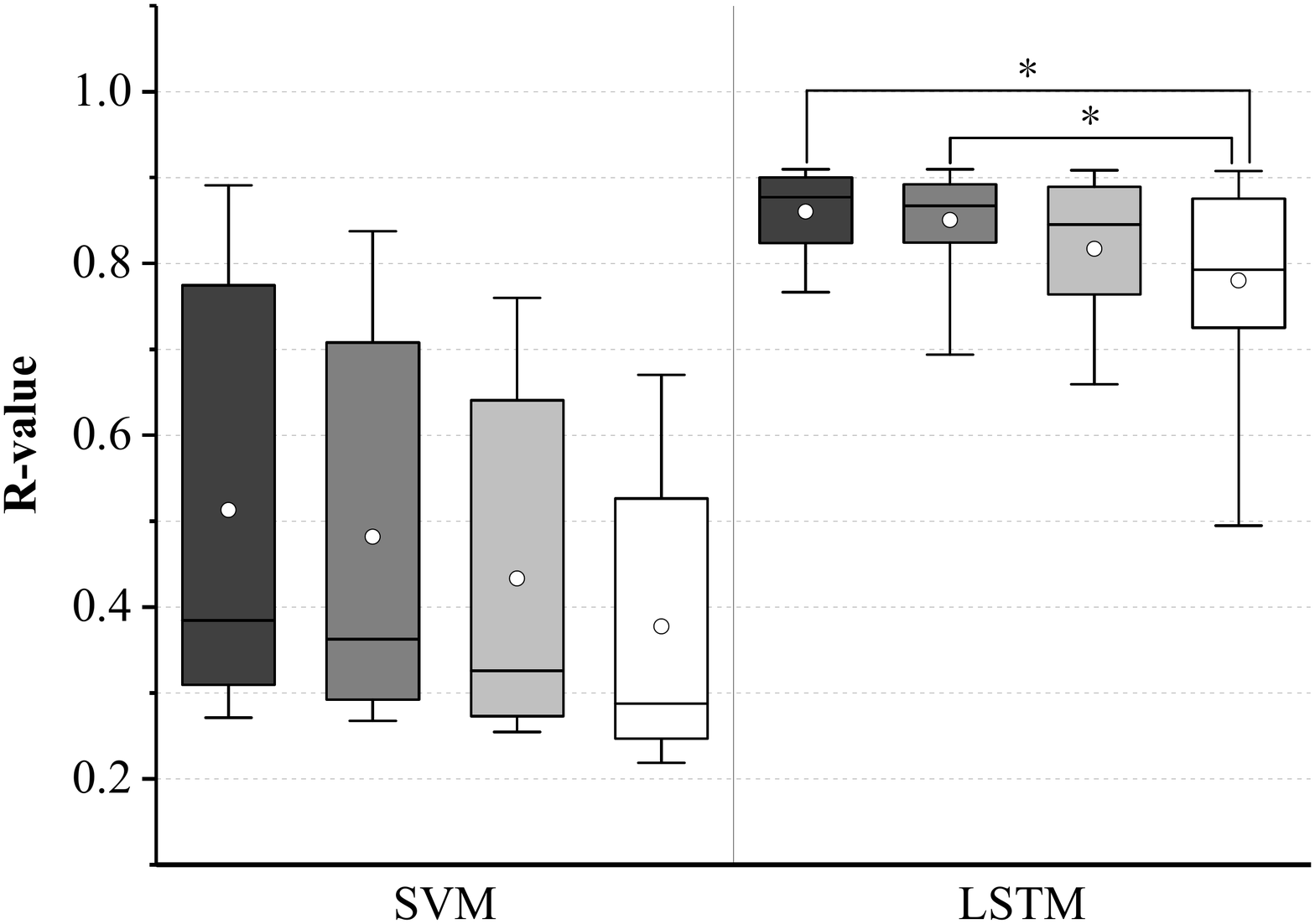}}
\caption{(a) RMSE, (b) SNR and (c) R-value of the performance of different prediction time over ten able-bodied subjects.  $*$ indicates a statistically significant difference (one-way ANOVA, P$<$0.05).} 
\label{fig:07}  
\end{minipage} 
\end{figure*}

\subsection{Evaluation of The Prediction Time}
As discussed above, six {time sections} are labeled as prediction
time to access the interval that could be used to compensate for
mechanical delay with the best performance. To do this, the Pearsons correlation coefficient (R-value),
the root mean square error (RMSE), the signal-to-noise ratio (SNR) and the adjusted R square are calculated
with different prediction time.

Fig. \ref{fig:061} presents the average performance of the six sections of prediction time. It can be seen from the figure that the accuracy of the last two time sections (135ms and 162ms) is obviously worse than the rest four sections of prediction time, regardless of the metrics. The relatively inferior performance of the last two time sections indicates the boundary of the prediction time.

Fig. \ref{fig:07} presents the comparison of different prediction time sections depicted by RMSE, SNR and R-value. Note that in this figure, only four time sections are presented.  Different grey levels denote different prediction time.  In Fig. \ref{fig:07}(a),
there is no significant difference of SVR predictor among all the prediction time (P$>$0.05). The best results are 12.99 deg for 27ms, 14.31  deg for 54ms, 14.34 deg for 81ms and 14.36 deg for 108ms,
respectively. Also,  the statistical analysis
shows there is no significant difference (P$>$0.05) among the RMSE of the LSTM predictor. The best
results of 27ms, 54ms, 81ms and 108ms are 0.66 deg, 0.69 deg, 1.48 deg, 1.36 deg, respectively.

As shown Fig. \ref{fig:07}(b), the SNRs of the SVM
predictor among different prediction time are statistically equal (P$>$0.05). The best
results are 2.49dB for 27ms, 2.52dB for 54ms,
2.52dB for 81ms and 2.43dB for 108ms, respectively. Averaged SNR of the LSTM predictor
generally decreases with the increment of  prediction time. The statistical analysis shows
there is no significant difference between 54ms and 108ms (P$>$0.05). The best results are 24.47dB for 27ms,
25.3dB for 54ms, 24.87dB for 81ms and 23.98dB for 108ms, respectively.

In Fig. \ref{fig:07}(c), the averaged R-value of both predictors presents a generally decreasing trend as the prediction time increases. No significant difference is found in the results of the SVM
predictor between 27ms and 108ms (P$>$0.05). And the best results are 0.89 for 27ms, 0.84 for 54ms, 0.76 for 81ms and 0.67 for 108ms,
respectively. For the LSTM predictor,
a significant difference exists between 27ms and 108ms
(P=0.009) and between 54ms and 108ms (P=0.0248). The
best results for the LSTM predictor are 0.91 for 27ms, 0.91 for 54ms, 0.908 for 81ms and 0.908 for 108ms, respectively.

It is in a statistic paradigm that the prediction time sections are set and the performance of different prediction time sections is evaluated. Accuracy improvement can be expected if  subject-specific EMD-inspired prediction time is employed for predicting joint angles. However, EMD
varies with muscle fatigue
and across the continuous repetitive motions. The intractable online detection of EMD time would make such a prediction time setting paradigm impossibly achievable for exoskeletons. Thus,  we employ time sections, rather than a specific value, in order to study the effects of EMD-inspired prediction time on angle prediction and to provide a quantified reference for the compensation of exoskeletons' transmission delay.  According to the performance of all the six time sections, the boundary of prediction time is experimentally explored. In addition, the prediction time influences the prediction accuracy. Compared with \cite{8886526} that predicted one time-step ahead, our RMSEs are slightly larger, which might be attributed to our significantly larger prediction time. And the performance of prediction time
between 27ms and 108ms does not show a significant difference and the performance of prediction time out of this interval presents obvious deterioration. 
This result indicates that any value in the time interval of 27ms and 108ms can be
determined as the prediction time.
The prediction time section for controlling exoskeletons
is recommended to  be initially set between 54ms and 81ms and tuned according to the predicting performance. The inference time of executing our  algorithm is 5-7ms, achieved by  Delsys Trigno SDK module and NVIDIA TX2
computing module. In this way, our prediction time is large enough to cover the execution of our algorithm and still leaves enough room for compensating the  mechanical transmission delay.

\subsection{Effectiveness of The Predictors}
For the purpose of evaluating
the performance of SVR and LSTM on the continuous kinematics
prediction, an accuracy comparison needs to be made between
the two predictors. We use R-value, SNR and RMSE to quantify the performance. 

\begin{figure*}[htbp] 
\centering 
\begin{minipage}[c]{0.33\textwidth} 
\centerline{\includegraphics[height=5.5cm]{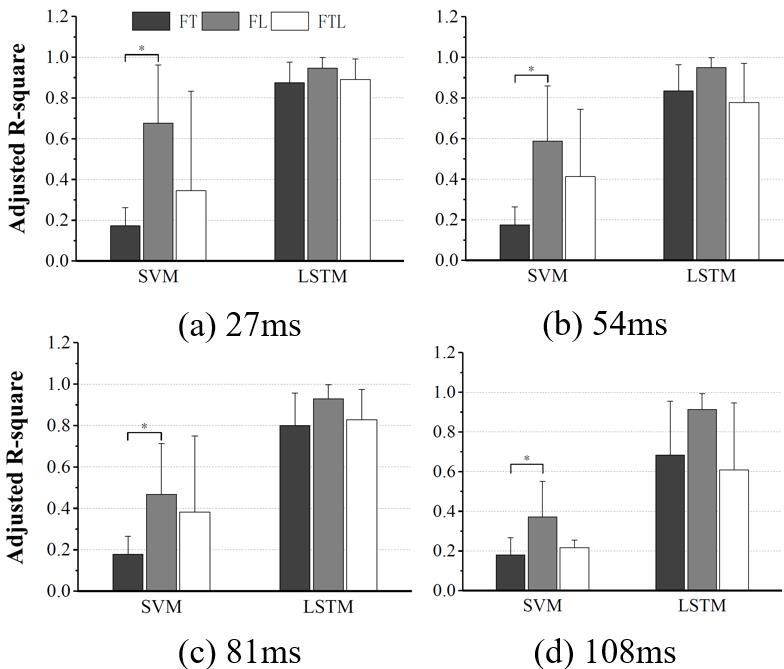}}
\caption{Statistic adjusted R-square results averaged over ten able-bodied subjects.
The adjusted R-square results refer to all designed EMG features using the
SVM and LSTM predictors were shown for individual prediction times: (a)
27ms, (b) 54ms, (c) 81ms, (d) 108ms. $*$ indicates a statistically significant
difference (one-way ANOVA, P$<$0.05). FT, FL and FTL denote EMG features
from time domain, EMG features from LSTM and EMG features from time
domain and LSTM, respectively.} 
\label{fig:09} 
\end{minipage} 
\hfill 
\begin{minipage}[i]{0.65\textwidth} 
\centering 
\centerline{\includegraphics[height=3.1cm]{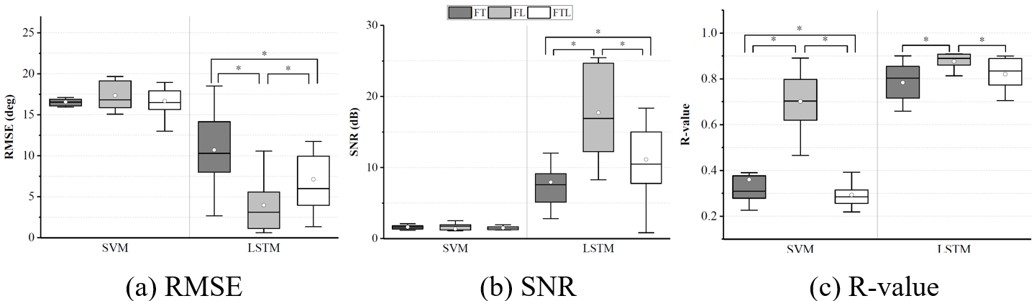}}
\caption{Statistic results of prediction comparison based on different features over ten able-bodied subjects.$*$ indicates a statistically significant difference (one-way ANOVA, P$<$0.05). (a) FT, (b) FL and (c) FTL denote EMG features from time domain, EMG features from LSTM and EMG features from time domain and LSTM, respectively.} 
\label{fig:08}   
\qquad
\centerline{\includegraphics[height=3.5cm]{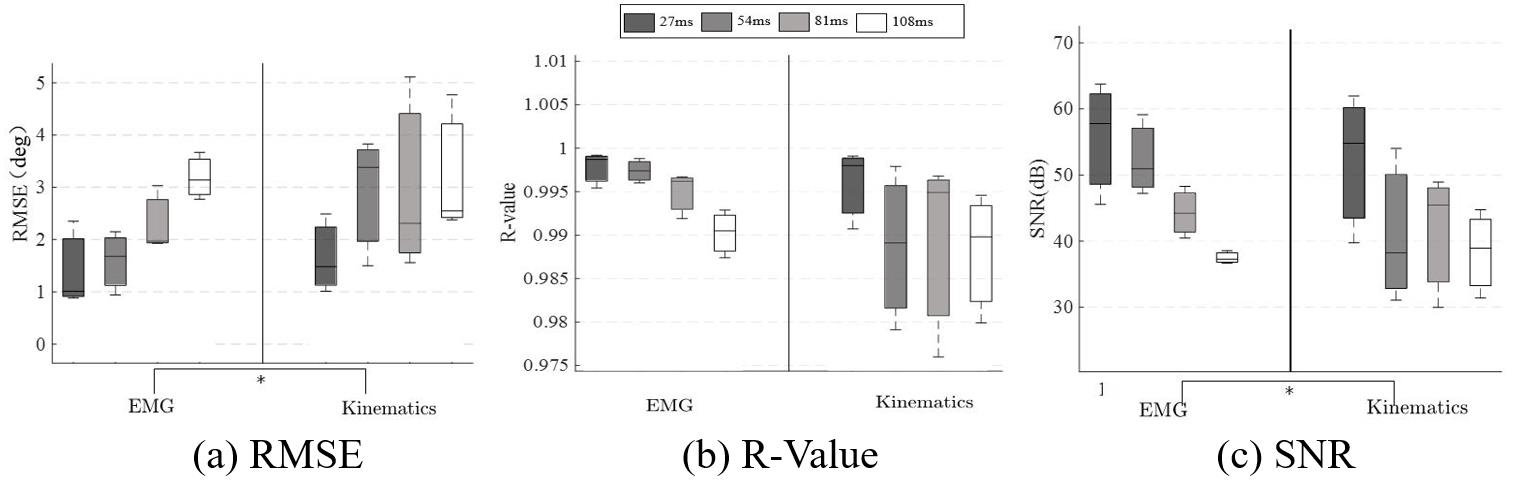}}
\caption{Statistic results of prediction comparison based on soly kinematics signals and fusion of kinematics and EMG signals over ten able-bodied subjects. $*$ indicates a statistically significant difference (one-way ANOVA, P$<$0.05). (a) RMSE, (b) R-value and (c) SNR denote different indexes for evaluation.} 
\label{fig:10}
\end{minipage}
\end{figure*}

As presented in Fig. \ref{fig:07}, the overall performance of
the predictors depicted by all the three indexes
present that the LSTM predictor significantly outperforms
the SVR predictor. (P$<$0.001).  It is also shown in  Fig. \ref{fig:09} and Fig.\ref{fig:08} that all the four indexes  of the LSTM predictor are obviously better than those of the SVM predictor.  It should be noted that the relatively larger RMSE of LSTM shown in Fig. \ref{fig:07} is due to the averaging over all the feature sets. It can be seen in Fig. \ref{fig:08} (a) that the LSTM predictor with the FL feature set is of good accuracy, with the RMSE of 3.98 deg.

The devised prediction method, which conducts an end-to-end kinematics prediction network (KinPreNet), achieves the best accuracy of predicting knee angles.  Moreover, the accuracy of the KinPreNet is better than that of the IMU-based angle calculation method we use \cite{yi2018estimating}, which indicates the predictor's ability of improving the errors of the input. And our better results, compared with the results of RCNN for predicting upper-limb motions \cite{xia2018emg}, suggest the benefit of our architecture. Based on the results, it can be concluded that LSTM
is more suitable for making a continuous and ahead-of-time prediction of knee angles. LSTM model benefits from its capability of learning to extract information deeply hidden in the input features through the regression process. Deep hierarchical representations of input features can be explored
by the devised LSTM predictor. The SVR predictor, with the lesser capability of exploring the hidden information of features,  could not explore the unobvious characters of the biomedical signals for this study.

\subsection{Validity of The Extracted Feature Sets}
To validate the completeness
of information incorporated in each feature set and their
consequent performance on angle prediction, R value, RMSE and SNR are estimated to present prediction accuracy. In addition, the adjusted R square is employed to evaluate
to what degree the accuracy variation of predicted angles is
correlated with different feature sets.

The average adjusted R-square for each prediction time
of ten subjects is shown in Fig. \ref{fig:09}. The performance of FL
outperforms the performance of FT and FTL. FL significantly
outperforms FT with all prediction time in the results of the SVR
predictor (P$<$0.05). Although the adjusted R square of FL is
shown to be better than FTL in both predictors, the difference
of FL and FTL are not statistically significant (P$>$0.05). The
FL-based prediction achieves better performance excepted
for the prediction from prediction time of 54ms and 108ms
with the LSTM predictor. In addition, with FL as the extracted
feature, the average adjusted R-square of the LSTM predictor is 27$\%$ - 54$\%$ higher than that of the SVR predictor. Statistical comparison of adjusted R-square among prediction time shows
that significant difference is not found when prediction time
changes for any EMG feature set (P$>$0.05).

Fig.\ref{fig:08} depicts the performance with different EMG feature sets.
The results of different EMG feature sets are represented by
different grey levels. In Fig. \ref{fig:08}(a), averaged RMSE of SVR
predictor are performed with three different EMG feature sets
(FL $>$ FTL $>$ FT). The best results of FT, FL and FTL are
15.95 deg, 15.07 deg and 12.99 deg, respectively. No significant difference
is found among features ( P(FT, FL)=0.077 ). Averaged results
with the LSTM predictor with FT, FL and FTL are 10.71$\pm$4.30 deg,
3.98$\pm$3.26 deg and 7.13$\pm$4.60 deg, respectively. Significant differences
are found in the results of the LSTM predictor among
all the EMG feature sets (P$<$0.01).

In Fig. \ref{fig:08}(b), SNR of the LSTM predictor significantly outperforms that of
the SVR predictor. There is no significant difference in the results
among different EMG feature sets of the SVR predictor (P$>$0.05).
The best results of FT, FL and FTL are 2.09dB, 2.52dB
and 1.95dB, respectively. Average SNR of the LSTM predictor
are depicted with different EMG feature sets (FL $>$ FTL $>$
FT). Significant differences are found in the results of the LSTM
predictor among all the features (P$<$0.01).

In Fig.  \ref{fig:08}(c), there are obviously significant differences
existed in R-value of the SVM predictor between FT $\&$ FL and
FL $\&$  FTL (P$<$0.001). Significant differences also exist in pairs
of FT and FTL (P=0.0277). The average R-value of FL can
up to 0.7$\pm$0.13 which is about 34$\%$ - 40$\%$ higher than that
of other EMG feature sets. Similar to SVR predictor, significant
differences are found in R-value results of the LSTM predictor
between FT $\&$ FL and FL $\&$ FTL (P$<$0.005). The average
R-value of FL with LSTM predictor is 0.88$\pm$0.04.

The LSTM feature extractor is established with the help of the
LSTM predictor. Obtaining model parameters of the LSTM
prediction system in the training session, the former part of
this system, defined as the LSTM extractor, can be used to
extract features from EMG signals. As shown in Fig. \ref{fig:08}, the
R-value, RMSE and SNR of FL outperform those of FT and
FLT with significant differences, which validates the effectiveness
of the LSTM extractor on prediction accuracy. The adjusted R square
value of FL demonstrates the LSTM extractor's
high correlation with the performance, which represents its
capability of explaining knee angle variations. In addition, no
significant difference exists among different sets of predictors
and prediction time, indicating the stability of the LSTM extractor.

The reason of such outstanding performance is twofold. Firstly, the joint training of the LSTM extractor and predictor, which propagates the error of angle prediction back to the
extractor during the training session, reinforces the correlation
between the extracted EMG features and knee angles. The
reinforced correlation contributes to well-decoded information
from multiple EMG channels. The results of \cite{ngeo2014continuous}, without presenting the capability of the ahead-of-time prediction, presented relatively lower accuracy, which also gives a side proof of the benefit of joint training. Secondly, the LSTM extractor
explores deep information inside the multi-channel EMG signals.
LSTM, developed for processing sequence information
with multi-layer neuron cells, is proper to decode the continuity
correlation between non-static EMG signals and kinematics
under noisy measurements. Meanwhile, rather than extracting
features from every single channel, the LSTM extractor extracts
features from multiple channels simultaneously contributing
to a comprehensive metric among muscles, which is related
to muscles antagonism during level walking.

Surprisingly, the performance of FTL is presented to be just
slightly better than that of FT, while no improvement is
observed comparing with the performance of FL. FT, representing
the overall information of the signal in a sliding analysis window,    
is just general time-domain features of EMG signals. In
contrast, features extracted from the LSTM extractor, due to
the function of the remember gate and forget gate in the LSTM model,
magnify the effect of dynamic components and minify the
effect of common resting components of EMG. Thus, FTL,
regarded as FL corrupted by FT, still contributes to a better
prediction than FT. Each FT is extracted from a single channel
of EMG electrodes, the effectiveness of which does not depend on FT from other channels. On the contrary, FL that is extracted
with joint information of multi-channel EMG signals works
as a whole in predicting knee angles. Hence, if compared with
FL, mixing FT with FL brings some redundancy information
into the extracted features, which contributes to a worsened
performance. But compared with FT, such a mixture adds some
deep characters into features, which improves the accuracy of
the prediction.

\subsection{Effectiveness of Employing EMG Signals}

The ahead-of-motion prediction  might result from two factors: the pseudo-periodic characteristics of gait and the electromechanical delay. In order to distinguish the influence of employing EMG signals, a comparison between predictions solely from kinematic signals and the fusion of kinematic and EMG signals is performed.   Due to the outstanding performance of FL combining with the LSTM predictor, the comparison under other sets of the feature extractors and  predictors is with little value. Thus, the comparison  is just performed using FL and the LSTM predictor. 

Fig. \ref{fig:10} presents the prediction performance with different prediction time using solely kinematics signals and the fusion of EMG and kinematics signals. There are significant differences existing in RMSE and SNR between the prediction from different signals, regardless of prediction time (P $<$ 0.05). And it can be seen from Fig.\ref{fig:10} that the performance of the prediction from EMG and kinematics signals generally outperforms that from solely kinematics signals.  

The performance comparison between prediction from kinematics signal and the fusion of EMG and kinematics signals demonstrates the benefits of employing EMG signals in the regression-based motion predicting. It can be concluded that although the pseudo-periodic characteristics of gait could make a contribution to the ahead-of-motion prediction, employing EMG signals plays a  necessary role given the  significantly improved accuracy. This phenomenon meets the significantly smaller prediction time reported in \cite{8886526} that solely exploited the pseudo-periodic characteristics of gait. Two factors might result in performance improvement. Firstly,  EMG signals from antagonistic muscle pairs around the knee relate to the moment and angular acceleration of the knee, which  incorporate vital information of knee motion changes. Thus, the intra-gait knee angle changes, which can hardly be predicted by the periodic gait characteristics, could be covered by features from EMG signals. Secondly, EMD provides ahead-of-motion information, which could consequently improve the prediction accuracy.

\section*{Acknowledgment}

This work is partly funded by National Natural Science Foundation of China (No. 61872112) and National Key Research and
Development Program of China (Nos. 2018YFC0806802
and 2018YFC0832105).





%



\bibliographystyle{IEEEtran}
\bibliography{./IEEEexample}
%








\end{document}